\documentclass{amsproc}
\usepackage{amsmath,amsthm, amscd,fancyhdr,graphics,pifont}
\usepackage{floatflt,subfigure,epsfig,moreverb,multicol,lscape}
\usepackage{supertabular,tabularx,multirow,bar,amssymb}
\usepackage{psfrag,afterpage,curves,epic, bbm}
\usepackage[english]{babel}

\newtheorem{theorem}{Theorem}[section]
\newtheorem{lemma}[theorem]{Lemma}
\newtheorem{corollary}[theorem]{Corollary}
\theoremstyle{definition}

\newtheorem{example}[theorem]{Example}
\theoremstyle{remark}
\newtheorem{remark}[theorem]{Remark}
\numberwithin{equation}{section}

\newcommand{\hvx}{\vect{\hat x}}

\newcommand{\supp}{\operatorname{supp}}

\newcommand{\matr}[1]{\mathsf{#1}}
\newcommand{\vect}[1]{\mathbf{#1}}
\newcommand{\code}[1]{\mathcal{#1}}
\newcommand{\set}[1]{\mathcal{#1}}

\newcommand{\GF}[1]{\mathbb{F}_{#1}}
\newcommand{\R}{\mathbb{R}}

\newcommand{\tr}{\mathsf{T}}
\newcommand{\convhull}{\operatorname{conv}}

\newcommand{\PG}[2]{\mathrm{PG}(#1,#2)}

\newcommand{\EGq}{\operatorname{EG}(2,q)}
\newcommand{\PGq}{\operatorname{PG}(2,q)}
\newcommand{\codePGq}{\code{C}_{\PGq}}

\newcommand{\defeq}{\triangleq}
\newcommand{\setDML}[1]{\set{D}^{\mathrm{ML}}_{#1}}
\newcommand{\setDMLzero}{\set{D}^{\mathrm{ML}}_{\vect{0}}}
\newcommand{\setDLPzero}{\set{D}^{\mathrm{LP}}_{\vect{0}}}
\newcommand{\vlambda}{\boldsymbol{\lambda}}
\newcommand{\vomega}{\boldsymbol{\omega}}

\newcommand{\vx}{\vect{x}}
\newcommand{\vy}{\vect{y}}

\newcommand{\edefinition}{\hfill$\square$}

\newcommand{\fp}[1]{\set{#1}}
\newcommand{\fph}[2]{\set{#1}(\matr{#2})}

\newcommand{\fch}[2]{\set{#1}(\matr{#2})}
\newcommand{\Mps}{\set{M}_{\mathrm{p}}}

\newcommand{\wps}{w_{\mathrm{p}}}
\newcommand{\wpsAWGNC}{w_{\mathrm{p}}^{\mathrm{AWGNC}}}

\newcommand{\wpsmin}{w_{\mathrm{p}}^{\mathrm{min}}}

\newcommand{\onenorm}[1]{\lVert #1 \rVert_1}
\newcommand{\twonorm}[1]{\lVert #1 \rVert_2}
\newcommand{\wH}{w_{\mathrm{H}}}

\newcommand{\pderiv}[2]{\frac{\partial #1}{\partial #2}} 

\begin{document}

\title[Bounds on the Pseudo-Weight of the Minimal Pseudo-Codewords]
      {Bounds on the Pseudo-Weight of \\
       Minimal Pseudo-Codewords of Projective Geometry Codes}

\author{Roxana Smarandache}\address{Department of Mathematics, San Diego State
  University, 5500 Campanile Dr., San Diego, CA 92182} 
\email{rsmarand@nd.edu}
\curraddr{Roxana
    Smarandache, Department of Mathematics, University of Notre Dame, Notre
    Dame, IN 46556}
\thanks{The first author would like to acknowledge the
support of NSF grant ITR-0205310}

  \author{Marcel Wauer} \address{ Department of Mathematics, San Diego State
    University, 5500 Campanile Dr., San Diego, CA 92182}
    \email{marcelwauer@gmx.de}

\date{October 1, 2005}


\begin{abstract}
  In this paper we focus our attention on a family of finite geometry codes,
  called type-I projective geometry low-density parity-check (PG-LDPC) codes,
  that are constructed based on the projective planes $\PGq$. In particular,
  we study their minimal codewords and pseudo-codewords, as it is known that
  these vectors characterize completely the code performance under
  maximum-likelihood decoding and linear programming decoding, respectively.
  The main results of this paper consist of upper and lower bounds on the
  pseudo-weight of the minimal pseudo-codewords of type-I PG-LDPC codes.
\end{abstract}

\maketitle

\section{Introduction}
The family of type-I PG-LDPC codes that we study here are cyclic codes which
moreover have very concise descriptions and large automorphism groups. It has
been observed experimentally that these codes have a good performance, close
to maximum-likelihood decoding, if decoded with iterative decoding or linear
programming decoding, see e.g.~\cite{Lucas:Fossorier:Kou:Lin:00:1,
  Kou:Lin:Fossorier:01:1,
  Vontobel:Smarandache:Kiyavash:Teutsch:Vukobratovic:05:1,Vontobel:Smarandache:05:1}.
Hence these codes are worthwhile study objects and this paper aims at
quantifying the difference between maximum-likelihood decoding and linear
programming decoding of these codes. 

The codes under consideration are defined as follows. Let $q \defeq 2^s$ for
some positive integer $s$, and consider a finite projective plane $\PGq$ (see
e.g.~\cite{Batten:97}) with $q^2 + q + 1$ points and $q^2 + q + 1$ lines: each
point lies on $q+1$ lines and each line contains $q+1$ points. A standard way
of associating a parity-check matrix $\matr{H}$ of a binary linear code to a
finite geometry is to let the set of points correspond to the columns of
$\matr{H}$, to let the set of lines correspond to the rows of $\matr{H}$, and
finally to define the entries of $\matr{H}$ according to the incidence
structure of the finite geometry. In this way, we can associate to the
projective plane $\PGq$ the code $\codePGq$ with parity-check matrix $\matr{H}
\defeq \matr{H}_{\PGq}$, whose parameters are $[q^2 + q + 1, n - 3^s - 1, q +
2].$ This code has the nice property that, with an appropriate ordering of the
columns and rows, the parity-check matrix is a circulant matrix, meaning that
$\codePGq$ is a cyclic code. 

\begin{example}
  \label{ex:pg:2:2}

  As an example, we take $q=2$ to obtain the Fano
  plane as in Figure~\ref{Fano} with a labeling that leads to a $7 \times 7$
  circulant matrix $\matr{H}_{\PG{2}{2}}$, and hence to a $ [7,3,4]$ binary
  cyclic code as its nullspace:
  \begin{align*}  \matr{H}_{\PG{2}{2}} &= 
      \begin{pmatrix} 
          1 & 1 & 0 & 1 & 0 & 0 & 0 \\
          0 & 1 & 1 & 0 & 1 & 0 & 0 \\
          0 & 0 & 1 & 1 & 0 & 1 & 0 \\
          0 & 0 & 0 & 1 & 1 & 0 & 1 \\
          1 & 0 & 0 & 0 & 1 & 1 & 0 \\
          0 & 1 & 0 & 0 & 0 & 1 & 1 \\
          1 & 0 & 1 & 0 & 0 & 0 & 1
              \end{pmatrix}.
  \end{align*}  
  \begin{figure}\label{Fano}
  \caption{$\PG{2}{2}$, with a labeling leading to a cyclic code}
  \begin{center}
     \epsfig{file=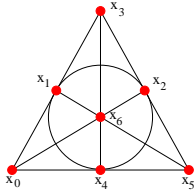, width=0.2\linewidth}
    \end{center} 
  \end{figure}
\end{example}

The paper is structured as follows. In Section \ref{definitions} we introduce
the framework necessary for the development of our results. We will talk about
maximum likelihood (ML) and linear programming (LP) decoding, and introduce
our main objects of study, pseudo-codewords, which are vectors that appear in
connection to iterative and linear program decoding.  Upper and lower bounds
on the pseudo-weight of minimal pseudo-codewords will be given in
Section~\ref{results}, which contains the new results of this paper.  In the
last section we offer some conclusions.

\section{Background and Definitions}\label{definitions}        
Consider a binary linear code $\code{C}$ of length $n$ and dimension $k$ that 
is used for data communication over a memoryless binary-input channel. The
codeword that is transmitted will be called $\vx$ whereas the received vector
will be called $\vy$. Based on the received vector, if we define the
log-likelihood ratios (LLR) to be $\lambda_i \defeq \log \big(
p_{Y_i|X_i}(y_i|0) / p_{Y_i|X_i}(y_i|1) \big)$, $i = 1, \ldots, n$, we can
view ML decoding as the optimization problem
\begin{align}
  \hvx
    &\defeq 
       \arg\min_{\vx \in \code{C}}
         \sum_{i=1}^{n}
           x_i \lambda_i=  
       \arg \min_{\vx \in \convhull(\code{C})}
         \sum_{i=1}^{n}
           x_i \lambda_i, 
             \label{eq:ml:decoder:2}
\end{align}
where $\convhull(\code{C})$ is the convex hull of $\code{C}$ in $\R^n$,
see~\cite{Vontobel:Smarandache:Kiyavash:Teutsch:Vukobratovic:05:1,
Vontobel:Smarandache:05:1, Feldman:03:1, Feldman:Wainwright:Karger:05:1}.
        
Among the codewords, the significant ones for ML decoding turn out to be the
{\em minimal} ones, $\set{M}(\code{C})$, i.e. the ones with support not
containing the support of any other non-zero codeword as proper subset.
Indeed, since we use a binary linear code $\code{C}$ over a binary-input
output-symmetric channel, we can without loss of generality
assume that the zero codeword was sent, because all decision regions are
congruent. Then we have that, if $\vx \in \code{C}$, the region
$\setDML{\vect{x}}$ in the LLR space where the ML decoder decides in favor of
the codeword $\vx$, $\setDML{\vect{x}}\defeq \big\{ \big.  \vlambda \in \R^n \
\big| \ \vx' \cdot \vlambda^\tr \geq \vx \cdot \vlambda^\tr \text{ for all }
\vx' \in \code{C} \setminus \{ \vx \} \big\}$,\footnote{Note that during ML
decoding, ties between decoding regions can either be resolved in a random or
in a systematic fashion.} shares a facet with the decision region
$\setDMLzero$ of the zero codeword if and only if $\vx \in
\set{M}(\code{C})$. Hence the set of minimal codewords are of particular
interest and are worth studying, as their patterns characterize the behavior
of the code under ML decoding.
 
For most codes of interest, the description complexity of
$\convhull(\code{C})$ grows exponentially
in the block length, therefore finding the minimum in \eqref{eq:ml:decoder:2}
with a linear programming solver is highly impractical for reasonably long
codes. Hence we relax the problem by replacing the above minimization by a
minimization over some easily describable polytope $\fp{P}$ that is a
relaxation of $\convhull(\code{C})$, i.e.
\begin{align}
   \hvx
    & \defeq \arg \min_{\vx \in \fp{P}}
        \sum_{i=1}^{n}
          x_i \lambda_i.
            \label{eq:lp:decoder:1}
 \end{align}
A relaxation commonly used, see~\cite{Feldman:03:1,
  Feldman:Wainwright:Karger:05:1, Koetter:Vontobel:03:1}, 
is given by {\em the fundamental polytope} $\fp{P}$ defined as
\vspace{-0.3cm}
\begin{align*}
  \fp{P}
    &\defeq
       \bigcap_{i=1}^{m}
         \convhull(\code{C}_i)
  \quad
  \text{ with }
  \quad
  \code{C}_i
     \defeq \left\{
              \vx \in \{0, 1\}^n
              \, \left| \, 
                \vect{r}_i \vx^\tr = 0
                \, \operatorname{mod}\, 2
              \right.
            \right\},
\end{align*}
where $\vect{r}_1, \vect{r}_2, \ldots, \vect{r}_m$ are the rows of a parity-check
matrix $\matr{H}$. Points in the set $\fp{P}$ are called {\em
  pseudo-codewords}. Because the set $\fp{P}$ is usually strictly larger than
$\convhull(\code{C})$, it can obviously happen that the decoding rule in
\eqref{eq:lp:decoder:1} delivers a vertex of $\fp{P}$ that is not a codeword.
Such vertices that correspond to pseudo-codewords that are not codewords are
the reason for the sub-optimality of LP decoding
(cf.~\cite{Feldman:Wainwright:Karger:05:1,Koetter:Vontobel:03:1}). Note that
$\fp{P} = \fph{P}{H}$ is a function of the parity-check matrix $\matr{H}$ that
describes the code $\code{C}$; different parity-check matrices for the same
code might therefore lead to different fundamental polytopes.

For LP decoding of a binary linear code that is used for data transmission
over a binary-input output-symmetric channel, it is sufficient to consider the
{\em fundamental cone} $\fch{K}{H}$, which is the part of the fundamental
polytope $\fp{P}$ around the vertex $\vect{0}$ and stretched to infinity. This
fundamental cone can be characterized as follows.

\begin{lemma}[\cite{Feldman:03:1, Feldman:Wainwright:Karger:05:1,
                         Koetter:Vontobel:03:1}]
  \label{lemma:fundamental:cone:1}
  
  Let $\code{C}$ be an arbitrary binary linear code and let $\matr{H}$ be its
  $m \times n$ parity-check matrix. Let $\vect{r}_1, \ldots, \vect{r}_m$ be
  the row vectors of $\matr{H}$. For each row vector $\vect{r}_j$,
  $j=\overline{1,m}$, we let $\supp(\vect{r}_j)$ denote its support in $\{1,
  \ldots, n\}$.

  Then, the fundamental cone $\fch{K}{H}$ is the set of vectors $\vomega \in
  \R^n$ with $\omega_i \geq 0, i=\overline{1,n}$, such that for each $j\in
  \{1,\ldots, m\}$ and for each $i \in \supp(\vect{r}_j)$,
  
 $$    \sum_{i' \in \supp(\vect{r}_j)
      \setminus \{ i \}} \omega_{i'} \geq \omega_{i}.$$
    \ 
 
\end{lemma}
Any point in $\fch{K}{H}$ will also be called a pseudo-codeword. We note that
if $\vomega \in \fch{K}{H}$, then also $\alpha \cdot \vomega \in \fch{K}{H}$
for any $\alpha > 0$, and so we will say that two such pseudo-codewords are in
the same equivalence class. Moreover, for any $\vomega \in \fch{K}{H}$ there
exists an $\alpha > 0$ (in fact, a whole interval of $\alpha$'s) such that
$\alpha \cdot \vomega \in \fph{P}{H}$.  Clearly, pseudo-codewords are not
codewords in general.
As before, if we let $\setDLPzero \defeq \big\{ \big. \vlambda \in \R^n \
\big| \ \vomega \cdot \vlambda^\tr \geq 0 \text{ for all } \vomega \in
\fph{P}{H} \setminus \{ \vect{0} \} \big\}$ be the region where the LP decoder
decides in favor of the codeword $\vect{0}$, it can easily been seen that
$\setDLPzero = \big\{ \big. \vlambda \in \R^n \ \big| \ \vomega \cdot
\vlambda^\tr \geq 0 \text{ for all } \vomega \in \fch{K}{H}
\big\}$.\footnote{Note that during LP decoding, ties between decoding regions
can either be resolved in a random or in a systematic fashion.} Moreover, we
have that $\setDLPzero = \big\{ \big. \vlambda \in \R^n \ \big| \ \vomega
\cdot \vlambda^\tr \geq 0 \text{ for all } \vomega \in \Mps(\fch{K}{H})
\big\}$ where $\Mps(\fch{K}{H})$ is the set of all vectors $\vomega \in
\fch{K}{H}$ such that the set $\{ \alpha \cdot \vomega \ | \ \alpha \geq 0 \}$
is an edge of $\fch{K}{H}$. Mimicking the parallel ML discussion above, we
will call these {\em minimal pseudo-codewords}.  Therefore, the set
$\Mps(\fch{K}{H})$ completely characterizes the behavior of the LP decoder.

\begin{example}
  Continuing Ex.~\ref{ex:pg:2:2}, the fundamental cone for $\matr{H} =
  \matr{H}_{\PG{2}{2}}$ is given by the set 
  {\begin{align*} 
    \fch{K}{H}
       = \big\{
      &{\vomega \in \R^7 \ | ~  \omega_i\geq 0, ~i=\overline{1,7}, }\\
      &{ -\omega_0 + \omega_1 + \omega_3 \geq 0, \quad +\omega_0 - \omega_1 +
        \omega_3 \geq 0, \quad
        +\omega_0 + \omega_1 - \omega_3  \geq 0  }, \\
      &{ -\omega_1 + \omega_2 + \omega_4 \geq 0, \quad +\omega_1 - \omega_2 +
        \omega_4 \geq 0, \quad 
        +\omega_1 + \omega_2 - \omega_4  \geq 0, } \\
      &{ -\omega_2 + \omega_3 + \omega_5 \geq 0, \quad +\omega_2 - \omega_3 +
        \omega_5 \geq 0, \quad 
        +\omega_2 + \omega_3 - \omega_5  \geq 0, } \\
      &{ -\omega_3 + \omega_4 + \omega_6 \geq 0, \quad +\omega_3 - \omega_4 +
        \omega_6 \geq 0, \quad 
        +\omega_3 + \omega_4 - \omega_6  \geq 0, } \\
      &{-\omega_4 + \omega_5 + \omega_0 \geq 0, \quad +\omega_4 - \omega_5 +
        \omega_0 \geq 0, \quad 
        +\omega_4 + \omega_5 - \omega_0  \geq 0, } \\
      &{ -\omega_5 + \omega_6 + \omega_1 \geq 0, \quad +\omega_5 - \omega_6 +
        \omega_1 \geq 0, \quad 
        +\omega_5 + \omega_6 - \omega_1  \geq 0, } \\
      &{ -\omega_6 + \omega_0 + \omega_2 \geq 0, \quad +\omega_6 - \omega_0 +
        \omega_2 \geq 0, \quad 
        +\omega_6 + \omega_0 - \omega_2  \geq 0}
       \big\}.
  \end{align*}}
Among these pseudo-codewords, the minimal ones are the ones with the number of
inequalities satisfied with equality larger than or equal to $n-1 = 7-1$, and
the rank of the matrix formed by the coefficients of these equalities is required
to be $n-1 = 7-1$.
\end{example}

\begin{remark}
  One can show that all minimal pseudo-codewords can be scaled such that all
  entries are non-negative integers. A word a caution: a pseudo-codeword
  $\vomega$ reduced modulo 2 might not be a codeword. However, there exists
  obviously a constant $\alpha$ (any even number would work) such that
  $\alpha\vomega$ reduced modulo 2 is a codeword. Throughout this paper we 
  consider pseudo-codewords that have the smallest possible integer 
  entries among all pseudo-codewords $\alpha\vomega$ with $\alpha>0$, 
  such that they are codewords if reduced modulo 2.
\end{remark} 
  
In the following we will only consider the additive white Gaussian noise
channel (AWGNC). In this context, we will define an important parameter in LP
decoding: the pseudo-weight of a pseudo-codeword, which corresponds to the
Hamming weight of a codeword in ML decoding. Indeed, the significance of the
Hamming weight $\wH(\vx)$ of a minimal codeword $\vx$ for ML decoding is the
following: it can be shown that the squared Euclidean distance from the point
$+\vect{1}$ in signal space, corresponding to the codeword $\vect{0}$, to the
boundary plane $\big\{ \vlambda \in \R^n \ | \ \vx \cdot \vlambda^\tr = 0
\big\}$ of the decision region of $\vect{0}$ under ML decoding is $\wH(\vx)$.
Similarly, {\em the AWGNC pseudo-weight} is defined such that if $\vomega$ is
a minimal pseudo-codeword then the squared Euclidean distance from the point
$+\vect{1}$ in signal space, corresponding to the codeword $\vect{0}$, to the
boundary plane $\big\{ \vlambda \in \R^n \ | \ \vomega \cdot \vlambda^\tr = 0
\big\}$ of the decision region of $\vect{0}$ under LP decoding is
$\wpsAWGNC(\vomega)$.

\begin{lemma}[\cite{Forney:Koetter:Kschischang:Reznik:01:1, 
                    Koetter:Vontobel:03:1}]
  The AWGNC pseudo-weight of a pseudo-codeword $\vomega \in \fch{K}{H}$ is
  equal to $\wpsAWGNC(\vomega) = \onenorm{\vomega}^2 / \twonorm{\vomega}^2$,
  where $\onenorm{\vomega}$ and $\twonorm{\vomega}$ are the $1$-norm and
  $2$-norm of $\vomega$.\footnote{Note that for $\vx \in \{ 0, 1 \}^n$ we have
  $\wpsAWGNC(\vx) = \wH(\vx)$, where $\wH(\vx)$ is the Hamming weight of
  $\vx$.} \edefinition
\end{lemma}

For smoother notations we will use $\wps(\vomega)$ in place of
$\wpsAWGNC(\vomega)$. Obviously any two pseudo-codewords of the same
equivalence class have the same pseudo-weight. Also, the minimum pseudo-weight
is always upper bounded by the minimum distance of the code. Hence, in order
to assess the performance under LP decoding, we need to look at the minimum
pseudo-weight over the set $\set{S}$ of all minimal pseudo-codewords that are
not multiples of minimal codewords. In particular we look at the {\em
pseudo-weight spectrum gap} $g(\matr{H})$:
$$g(\matr{H}) \defeq \min_{\vomega \in \set{S}} \wps(\vomega) - d_{min}.$$
For a randomly constructed code, the pseudo-weight spectrum gap can be negative.
However for the $\PGq$- and $\EGq$-based codes, the pseudo-weight spectrum gap
is \emph{non-negative}, which reflects the good performance,
close to maximum-likelihood, of these codes, under LP decoding. In fact, for the codes we
investigated, the pseudo-weight spectrum gap is significantly positive.

\begin{example}
  In \cite{Vontobel:Smarandache:Kiyavash:Teutsch:Vukobratovic:05:1,
  Vontobel:Smarandache:05:1} we were able to classify the minimal codewords of
  projective and Euclidean plane codes, with small $q$. We show here the
  results for $q=2$ and $q=4$.  If $q=2$ we obtain a $[7,3,4]$ code with 8
  codewords, all of which are minimal.  These are: $ (0,0,1,0,1,1,1),$ $
  (1,0,0,1,0,1,1),$ $ (1,1,0,0,1,0,1), $ $ (1,1,1,0,0,1,0),$
  $(0,1,1,1,0,0,1),$ $ (1,0,1,1,1,0,0), $ $ (0,1,0,1,1,1,0).$ It can be swown
  that these minimal codewords together with the following list of
  pseudo-codewords are the only minimal pseudo-codewords of this code. The
  list of non-codewords minimal pseudo-codewords are: $(2,2,1,2,1,1,1),$
  $(1,2,2,1,2,1,1),$ $(1,1,2,2,1,2,1)$, $(1,1,1,2,2,1,2),$ $(2,1,1,1,2,2,1)$,
  $(1,2,1,1,1,2,2),$ $ (2,1,2,1,1,1,2).$ The AWGNC pseudo-weight of these
  pseudo-codewords is $6.25$, so we obtain a strictly positive gap of $2.25$.
  In the case of the projective plane over $\GF{q}$ where $q = 4$, we have a
  $[21, 11, 6]$ code with the set of minimal codewords given by all codewords
  of weight $6$, $8$, and $10$. Moreover, one can show that the gap is
  $g(\matr{H}_{\PG{2}{4}}) = 9.8 - 6 = 3.8$, again strictly
  positive.\footnote{This result was obtained by simplifying the problem by
  using symmetries and then doing some brute-force computations with the help
  of ``lrs''~\cite{Avis:00:1}.} The next code based on $\PG{2}{8}$ behaves
  similarly. For details we refer
  to~\cite{Vontobel:Smarandache:Kiyavash:Teutsch:Vukobratovic:05:1,
  Vontobel:Smarandache:05:1}.
\end{example}  


\section{Bounds on the Pseudo-Weight of Pseudo-Codewords}
\label{results}


In this section we present results on projective-plane-based codes and their
minimum pseudo-weights.  In the following, $q$ will be a power of 2, and
$n=q^2+q+1$.
\begin{lemma}\label{proppseudo}
  Let $\code{C}$ be a code from projective plane over $\GF{q}$, with
  $\matr{H}\defeq\matr{H}_{\PG{2}{q}}$ an $n\times n$ circulant parity-check
  matrix.  Let $\vect{\vomega}$ be a non-zero pseudo-codeword. For any $l\in
  \{1,2,\ldots,n\}$ we have
  $$\sum_{i=1\atop i\neq l}^n \omega_i\geq (q+1)\omega_l,$$
 or, equivalently, $$\sum_{i=1}^n \omega_i\geq (q+2)\omega_l.$$
\end{lemma}
\begin{proof} We use the notation of Lemma~\ref{lemma:fundamental:cone:1}.
  Let $P$ be a point in the projective plane $\PGq$ that has the associated
  pseudo-codeword component value $\omega_l$. There are $q+1$ lines
  $\vect{r}_{j_i}, i\in\{1,\ldots,q+1\}$ crossing $P$. For each line
  $\vect{r}_{j_i}$ among the equations characterizing the fundamental cone
  $\fch{K}{H}$, $i\in\{1,\ldots, q+1\}$, we find: $$\sum_{i' \in
  \supp(\vect{r}_{j_i}) \setminus \{ l \}} \omega_{i'} \geq \omega_{l}.$$
  Because the sets $\vect{r}_{j_i} \setminus \{ P \}$, $i\in\{1,\ldots,
  q+1\}$, partition the set of all points minus $P$, we obtain the desired
  inequalities by adding these $q+1$ equations.
\end{proof}

The following theorem gives upper and lower bounds on the pseudo-weight of all
minimal pseudo-codewords. The two bounds have been already derived by several
other authors, e.g.~\cite{Vontobel:Koetter:04:1}, however the proof we present
here makes use of some very simple arithmetic, and therefore we include it as
well.

\begin{theorem}
  \label{th:wps:upper:lower:bound:1}
  Let $\code{C}$ be a code from projective plane over $\GF{q}$ with
  $\matr{H}_{\PG{2}{q}}$ an $n\times n$ circulant parity-check
  matrix. Let $\vect{\vomega}$ be a nonzero
  pseudo-codeword. Then $$ q+2\leq \wps(\vect{\vomega})\leq
  |\supp(\vect{\vomega})|.$$  
\end{theorem}

\begin{proof} 
  We have that $$\sum_{l=1}^n \omega_l\sum_{{ i=1\atop i\neq l}}^n
  \omega_i=\onenorm{\vomega}^2-\twonorm{\vect{\vomega}}^2
  \quad\Rightarrow\quad
  \wps(\vect{\vomega})=1+ \frac{\sum\limits_{l=1}^n \omega_l\left(\sum\limits_{i=1\atop
        i\neq l}^n \omega_i\right)}{ \twonorm{\vect{\vomega}}^2}.$$
  Using Lemma\ref{proppseudo}, i.e. $\sum\limits_{l=i\atop i\neq l}^n \omega_i\geq
  (q+1)\omega_l$, we obtain the lower bound $ \wps(\vect{\vomega})\geq
  q+2$. The upper bound is obtained using the Cauchy-Schwarz
  inequality. Indeed, the desired upper bound follows easily from
  \begin{align*}
    \left(\sum_{i=1}^n \omega_i\right)^2
      &= \left(
           \sum_{i \in \supp(\vomega)}
             1 \cdot\omega_i\right)^2
       \leq
         \left(
           \sum_{i \in \supp(\vomega)} 1^2
         \right)
         \left(\sum_{i \in \supp(\vomega)}
           \omega_i^2
         \right)  \\
      &= |\supp(\vect{\vomega})|
         \cdot
         \left(
           \sum_{i \in \supp(\vomega)} \omega_i^2
         \right).
  \end{align*}
\end{proof}
 
Let now $\wpsmin(\matr{H})$ be the minimum AWGNC pseudo-weight of a code with
parity-check matrix $\matr{H}$. Since the above lower bound matches the
minimum Hamming weight, the following observation is immediate.
 
\begin{corollary}
  The minimum pseudo-weight of a projective plane code is $
  \wps^{min}(\vect{\vomega})=q+2=d_{min}.$
\end{corollary}


The upper bound in Th.~\ref{th:wps:upper:lower:bound:1} is obviously attained
for codewords. However, for minimal pseudo-codewords that are not multiples of
codewords, this upper bound seems to be the further away from the actual
weight, the larger the bound. We are interested in these vectors, especially
those with small pseudo-weight, because they are relevant in quantifying the
spectrum gap, which in turn quantifies the performance difference between LP
and ML decoding. Hence, we would like to give certain estimates on the possible
minimum pseudo-weight of such vectors. Our observation, made by studying the
behavior of vectors in certain examples, was that pseudo-codewords
of small support are candidates for small weight pseudo-codewords, and among
these, the ones with values of 0, 1, and 2 for the entries. In the remaining of
this section we will give upper and lower bounds on the pseudo-weight of
vectors with integer entries, and on minimal pseudo-codewords in particular,
with a special emphasis on the ones with entries equal to 0, 1, or 2.
  
\begin{theorem}[Conjecture]
  The smallest weight among all minimal pseudo-codewords that are not
  codewords is given by the pseudo-codewords that contain only zeros, ones, and
  twos.
\end{theorem}

\subsection{Minimization of the Pseudo-Weight of a Vector}\label{secmini}

We would like to mention that the results that we present in this subsection
were derived in by the second author in~\cite{Wauer:05:1}. The main result is
presented in Theorem~\ref{thlower} and consists of a lower bound on the
pseudo-weight of a vector. Since this theorem does not use any property of a
pseudo-codeword, it holds for any vector $\vect{x}$ that has non-negative
entries. In the following we will use the notation $\vx$ to identify an
arbitrary vector with non-negative entries and $\vomega$ to identify a
pseudo-codeword. Before presenting this theorem, we study first the influence
of $0$-components on the pseudo-weight.
 \begin{lemma}\label{lemma31}
   Let $\vx$ be a vector of length $n$ with with $r$ $0$-components and $n - r$
   positive entries $x_i$ where $r < n - 1$. Then there exists a vector $\vx'$ of
   length $n$ with $n - r - 1$ positive entries such that $\wps(\vx') <
   \wps(\vx)$.
 \end{lemma}
 \begin{proof} First we multiply $\vx$ by an $\alpha$ such that the smallest non-zero
 component in $\tilde{\vx} = \alpha \vx$ is $1$. The vector $\vx'$ is obtained
 by switching this $1$-component $\tilde{x}_l$ in $\tilde{\vx}$ to $0$. Let $s
 \defeq \sum\limits_{i=1\atop {i\neq l}}^n \tilde{x}_i$ and $t\defeq
 \sum\limits_{i=1\atop{i\neq l}}^n \tilde{x}_i^2$. From the assumptions in the
 lemma statement, it follows that $s\geq 1$ and $t\geq 1$. Moreover it holds
 that $ s\leq t$. Therefore $1+\frac{1}{t}\leq 1+\frac{1}{s}$ and because $
 1+\frac{1}{s}>1$ we have that $1+\frac{1}{t} < (1+\frac{1}{s})^2$ or
 equivalently, $\frac{s^2}{t} <\frac{(s + 1)^2}{t+1}$.  Then:
 \begin{equation*}
 \wps(\vx') 
 = \frac{s^2}{t} < \frac{(s + 1)^2}{t + 1} = \wps(\tilde{\vx}) = \wps(\vx).
 \end{equation*}
\end{proof}
To get prepared for the main result of this subsection we next show that the
minimal pseudo-weight is obtained for vectors with a certain structure, namely
vectors that contain only two different components. More precisely, we will
prove that among all non-zero vectors with non-negative integer components in
an interval $[m,M]$, $0<m\leq M$, and with at least one component equal to $m$
and one equal to $M$, the vectors with minimal pseudo-weight
are the ones that have the non-zero components equal to either $m$ or $M$.

\begin{lemma}\label{lemma45} Let $0<m\leq M$. Among all vectors 
  $\vx = (x_1, \ldots, x_n)$ with $x_j = 0$ or $m \leq x_j \leq M,~
  j\in\{1,\ldots,n\}$, and with at least one $x_{j_1}=m$ and one $x_{j_2}=M$,
  the vectors with minimal pseudo-weight are the ones that have all non-zero
  components equal to either $m$ or $M$.
\end{lemma}
\begin{proof} Let $\vx = (x_1, \ldots, x_n)$ be a vector with the properties
  of the lemma statement. Suppose that $\vx $ has a components $x_j$ with $m <
  x_j < M$. Let $A \defeq \sum\limits_{i=1,i\neq j}^nx_i$ and $B \defeq
  \sum\limits_{i=1,i\neq j}^nx_i^2$. The pseudo-weight of $\vx$ is
\begin{equation*}
\wps(\vx) = \frac{(A + x_j)^2}{B + x_j^2}.
\end{equation*}
The first and second partial derivative of the pseudo-weight as a function of
$x_j$ are, respectively,
\begin{align*}
  \pderiv{\wps}{x_j}
    &= -\frac{2(A + x_j)(Ax_j - B)}{(B + x_j^2)^2}, \\
  \frac{\partial^2 \wps}{\partial {x_j}^2 }
    &= \frac{2(B^2 - 3Bx_j^2 - 6ABx_j + 2Ax_j^3 + 3A^2x_j^2 - A^2B)}
             {(B + x_j^2)^3}.
\end{align*}
The critical points of the pseudo-weight function are therefore $x_{j,1} = -A
< 0$ and $x_{j,2} = \frac{B}{A} > 0$. With $\frac{\partial^2 \wps}{\partial
{x_j}^2 } (\frac{B}{A}) = -\frac{2A^4}{B^2(B + A^2)} < 0$ we obtain an
absolute maximum at $x_2 = \frac{B}{A}$, i.e.~the absolute minimum of the
pseudo-weight function over $[m,M]$ is at one of the endpoints of $[m,M]$.
Switching $x_j$ to either $m$ or $M$ generates now a vector with smaller
pseudo-weight. The claim follows by repeating this procedure for all $m < x_l
< M$.
\end{proof}
  We are
ready now for the main result of this subsection.  

\begin{theorem}
  \label{thlower}

  Let $\vx = (x_1,\ldots,x_n)$ be an arbitrary vector with non-negative
  integer components. Let $N \defeq |\supp(\vx)|$, and let $M$ and $m$ be the
  largest and smallest value among the non-zero entries of $\vx$,
  respectively. Then
  \begin{equation*}
\wps(\vx) \geq N \cdot \frac{4mM}{(m + M)^2},
\end{equation*}
with equality if and only if $\vx$ contains $N \cdot \frac{M}{m + M}$
components of value $m$, and $N \cdot \frac{m}{m + M}$ components of value $M$.
\end{theorem}
\begin{proof} According to Lemma \ref{lemma45} we only consider vectors with
  $t_m$ components equal to $m$ and $t_M$ components equal to $M$. With $t_M
  = N - t_m$, the pseudo-weight of $\vx$ is given as function $f$ of $t_m$:
\begin{equation*}
f(t_m)=\wps(\vx) =
 \frac{(mt_m + Mt_M)^2}{m^2t_m + M^2t_M} = 
\frac{(mt_m + M(N - t_m))^2}{m^2t_m + M^2(N -
 t_m)}.
\end{equation*}
Its first and second derivative are then given by, respectively,
\begin{align*}
  \frac{\partial f(t_m)}{\partial t_m}
    &= -\frac{((m - M)t_m + M N)((-(m + M)(M -
       m)^2)t_m + 
       MN(M - m)^2)}{(m^2t_m + M^2N - M^2t_m)^2}, \\
  \frac{\partial^2 f(t_m)}{\partial {t_m}^2}
    &= -\frac{2(mM(M - m)N)^2}{(m^2t_m
      + M^2N - M^2t_m)^3}.
\end{align*}
We see that the critical points of the function $f$ are at $t_{m,1} =
\frac{MN}{M - m}$ and $t_{m,2} = \frac{-MN(M - m)^2}{-(m + M)(M - m)^2} =
\frac{MN}{m + M}$, with second derivative given by $\frac{\partial^2
f}{\partial {t_m}^2 }(t_{m,1}) = -\frac{2(M - m)^2}{mMN} < 0$ and
$\frac{\partial^2 f}{\partial {t_m}^2 }(t_{m,2}) = \frac{2(M - m)^2}{mMN} >
0$, respectively. Hence the function $f$ has an absolute minimum at $t_{m,2}
=\frac{MN}{m + M}$ with value $f(t_{m,2}) = \frac{4mMN}{(m + M)^2}$. Finally,
we calculate $t_M = N - t_{m, 2} = \frac{mN}{m + M}$.
\end{proof}
\begin{corollary}
  \label{cor:wps:lower:bound:1}
  Let $\vx = (x_1,\ldots,x_n)$ be an arbitrary vector with non-negative
  integer components. Let $N \defeq |\supp(\vx)|$, let $M$ and $m$ be the
  largest and smallest value among the non-zero entries of $\vx$,
  respectively, and let $r\defeq \frac{M}{m}.$ Then
\begin{equation*}
\wps(\vx) \geq N \cdot \frac{4r}{(r + 1)^2},
\end{equation*}
with equality if and only if $\vx$ contains $N \cdot \frac{r}{r + 1}$
components of value $m$, and $N \cdot \frac{1}{r + 1}$ components of value $M$.
\end{corollary}

\begin{proof}
  This follows directly from Theorem \ref{thlower}.
\end{proof}

It is easy to construct pseudo-codewords whose pseudo-weight achieves the
lower bounds in Theorem~\ref{thlower} and Cor.~\ref{cor:wps:lower:bound:1},
e.g.~pseudo-codewords where all non-zero entries have the same value achieve
these lower bounds.

\begin{example}
  For vectors of length $N$ with $m = 1$ the corresponding lower bound $L$
  of the pseudo weight $\wps(\vx)$ is given in Table~\ref{tableEx}.
\begin{table}
\caption{Lower bounds on the pseudo-weight.}\label{tableEx}
\begin{center}
\begin{tabular}{|c| c| |c| c| |c| c|}    \hline
$M$ & $L$ & $M$ & $L$ & $M$ & $L$\\ \hline\hline
&&&&&\\
$1$ & $N$              & $6$  & $\frac{24}{49}N$  & $11$ & $\frac{11}{36}N$\\&&&&&\\
$2$ & $\frac{8}{9}N$   & $7$  & $\frac{7}{16}N$   & $12$ & $\frac{48}{169}N$\\&&&&&\\
$3$ & $\frac{3}{4}N$   & $8$  & $\frac{32}{81}N$  & $13$ & $\frac{13}{49}N$\\&&&&&\\
$4$ & $\frac{16}{25}N$ & $9$  & $\frac{9}{25}N$   & $14$ & $\frac{56}{225}N$\\&&&&&\\
$5$ & $\frac{5}{9}N$   & $10$ & $\frac{40}{121}N$ & $15$ &
$\frac{15}{64}N$\\&&&&&\\ 
\hline 
\end{tabular}
\end{center}
\end{table}
\end{example}


\subsection{Bounds on the Pseudo-Weight of Pseudo-Codewords}
We will now apply the general results obtained in the Section~\ref{secmini} to
pseudo-codewords. In particular we will give a lower bound on the
pseudo-weight of pseudo-codewords with entries equal to zero, one, and two
only, as these are the ones suspected to have the smallest pseudo-weight among
all \emph{minimal} pseudo-codewords that are not codewords. We will also look at the
pseudo-codewords of maximum support $n$, as these are candidates for having
the largest pseudo-weight among all \emph{minimal} pseudo-codewords that are not
codewords. In the following, we will use the combinatorics terminology of a
multiset $\{t_0\cdot 0, t_1\cdot1,\ldots, t_M\cdot M\}$ to display the values
taken by the components together with the number of times a value $i \in
\{0,1,\ldots, M\}$ is taken by the components of $\vomega$. Naturally,
$|\supp(\vomega)| = \sum\limits_{i=1}^{M}t_i$. We begin our considerations
with a lemma.
\begin{lemma}\label{support}  Let $\code{C}$ be a code from projective plane over $\GF{q}$, with
  $\matr{H}_{\PG{2}{q}}$ an $n\times n$ circulant parity-check
  matrix. Let $\vect{\vomega}$ be a nonzero
  pseudo-codeword  with the set of entries given by the multiset $\{t_0\cdot 0, t_1\cdot1,
  t_2\cdot 2\}$. Then: $$|\supp(\vomega)| \geq \frac{3}{2}(q+2).$$ 
\end{lemma}
\begin{proof}
Applying Lemma~\ref{proppseudo} we obtain $$\sum_{l=1}^2 l\cdot t_l\geq
2(q+2).$$
Since  $\vect{\vomega}$ reduced modulo 2 is a codeword, we
have that $t_1\geq q+2. $
Adding the last 2 equations yields $$2(t_1+t_2)\geq 3(q+2) \quad \Rightarrow
\quad 2|\supp(\vomega)|\geq
3(q+2).$$ The desired bound now follows.
  \end{proof}

\begin{corollary}  Let $\code{C}$ be a code from projective plane over $\GF{q}$ with
  $\matr{H}_{\PG{2}{q}}$ an $n\times n$ circulant parity-check
  matrix. Let $\vect{\vomega}$ be a nonzero
  pseudo-codeword  with the set of entries given by the multiset $\{t_0\cdot 0, t_1\cdot1,
  t_2\cdot 2\}$. Then the inequality holds
  \begin{align*}
    \wps(\vect{\vomega}) &\geq \frac{4}{3}(q+2).
  \end{align*}
\end{corollary}
\begin{proof} 
 Applying Lemma~\ref{support} together with Theorem~\ref{thlower} we obtain: 
 $$\wps(\vect{\vomega})\geq \frac{8}{9}\cdot\frac{3}{2}(q+2)=\frac{4}{3}(q+2).$$  

\end{proof}

\subsection{Maximization Of The Pseudo-Weight}


In this section we give a way of constructing a minimal pseudo-codeword of
largest possible support $n$, in a projective plane over $\GF{q}$, for any $q$
even. We conjecture that this has the largest pseudo-weight among all minimal
pseudo-codewords that are not codewords.

\begin{theorem} Let $\code{C}$ be a code from projective plane over $\GF{q}$, with
  $\matr{H}_{\PG{2}{q}}$ an $n\times n$ circulant parity-check matrix. Let
  $L=\{P_j~| ~j\in \{1, 2,\ldots, q+1\}\}$ be an arbitrary line in the
  projective plane, and let $\set{S}$ be its support. Let $\vect{\vomega}\in
  \R^n$ such that $\vect{\vomega}_i=q$ for all $i\in \set{S}$ and
  $\vect{\vomega}_i=1$ for all $i\in \{1,2,\ldots,n\}-\set{S}$.  Then
  $\vect{\vomega}$ is a minimal pseudo-codeword with pseudo-weight equal to
  $\frac{(2q+1)^2}{q+2}$.
\end{theorem}
\begin{remark}
This vector is obtain by assigning values of $q$ to the points on a line at infinity,
and assigning values of $1$ to the points of the remaining affine plane.  
\end{remark}
\begin{proof} 
Each point $P_i$ of associated value $\vomega_{i}=q$, belongs to $q$
other lines, all intersecting $L$ in exactly one point. Each of these $q$
lines has a set of $q+1$ points of associated values $q$ for the point $P_i$,
and $1$ for the remaining $q$ points. Among the fundamental cone inequalities
associated to the equation of line $L$, only one will be satisfied with
equality, $-q+1+1+\ldots+1=0$, all the other inequalities being satisfied
strictly. This is valid for each point on the line $L$, hence there are a
total number of $q(q-1)=q^2-q=n-1$ equations satisfied by $\vect{\vomega}$.
Thus far we obtained that $\vect{\vomega}$ is a pseudo-codeword that attains
$n-1$ equality among the defining inequalities of the fundamental cone.  We
need to show that the pseudo-codeword is minimal, i.e.~we need to show that
the matrix $\matr{A}$ such that $\matr{A}\cdot\vect{\vomega}=0$ over the real
field of numbers, has maximum possible rank $n-1$.  Computing the matrix
$\matr{A}\matr{A^T}$ results in the matrix
$$\matr{A}\matr{A^T}=q{\rm {\rm I}_{n-1}} +{\matr{1}_{n-1}}$$ with
$\matr{1}_{n-1}$ the matrix with all entries equal to 1. The eigenvalues of
the matrix $\matr{A}\matr{A^T}$ are given by $q$, (with multiplicity $n-2$)
and $q^2+2q$ (with multiplicity 1).

It implies that the matrix $\matr{A}\matr{A^T}$ is positive definite, hence,
$\matr{A^T}\vect{\vomega}=0$ if and only if $\vomega=0$.  It follows that the
rank of $\matr{A}$ is equal to $n-1$.
\end{proof}

\begin{theorem}[Conjecture]\label{bounds}
  If $\vect{\vomega}$ is a minimal pseudo-codeword that is not a codeword, 
 then $$\wps(\vect{\vomega})\leq
  \frac{(2q+1)^2}{q+2}.$$

\end{theorem}

\section{Conclusions}
In this paper we discussed some upper and lower bounds on the pseudo-weight of
pseudo-codewords of type-I PG-LDPC codes, and gave examples of
pseudo-codewords attaining the bounds. We also conjectured that the low
support, low entry integer minimal pseudo-codewords are the ones of minimal
pseudo-weight among the non-codeword minimal pseudo-codewords.

\section*{Acknowledgments} We would like to thank Dr.~Pascal O.~Vontobel for 
 our many valuable discussions during the research that led to this
 paper. 

\bibliographystyle{ieeetr}
\bibliography{../BibFiles/huge,../BibFiles/Pascal_ref}


\end{document}